\documentclass[letter, twocolumn]{jpsj2}

\newcommand{\nnpair}[1]{\left < #1 \right >_{\text{n.n.}}}
\newcommand{\ave}[1]{\langle #1 \rangle}

\title{Mean-Field Analysis of Electric Field Effect on Charge Orders in Organic Conductors}

\author{Emi \textsc{Yukawa}\thanks{E-mail address: yukawa@hosi.phys.s.u-tokyo.ac.jp} and Masao 
\textsc{Ogata}}

\inst{Department of Physics, University of Tokyo \\ Hongo, Bunkyo-ku, Tokyo, 113-0033 JAPAN}

\abst{In order to investigate charge ordering phenomena under electric field, static nonequilibrium 
Hartree approximation (SNHA) method is formulated on the basis of the nonequilibrium Green's 
functions introduced by Keldysh. By applying the SNHA to the 3/4-filling extended 
Hubbard model on anisotropic triangular lattice, we study the stabilities and amplitudes 
of 3-fold and horizontal charge orders in ${\theta}$ and 
${\theta}_d$-(BEDT-TTF)$_2X$ salts under the electric field. 
The obtained results show that the electric field stabilizes the 3-fold state in comparison to 
the horizontal state. The amplitude of the 3-fold state tends to decrease by the field, whereas that of 
the horizontal state does not change. }

\kword{charge order, anisotropic triangular lattice, organic conductor, 
static nonequilibrium condition, Hartree approximation, nonequilibrium Green's function}

\begin{document}
\maketitle
One of the families of organic crystals, (BEDT-TTF)$_2X$ (in short (ET)$_2X$), has quasi-two-
dimensional nature. Its donors, BEDT-TTF (ET) molecules, and acceptors, $X$s, form monolayers 
respectively, and they are stacked layer by layer. Spatial arrangements of ET molecules can be 
controlled by the anion molecules and hydrostatic pressure~\cite{3}. Among the 
different polytypes, ET layers in $\theta$ and ${\theta}_d$-type crystals, which are focused on in 
this paper, form 3/4-filling anisotropic triangular lattices: the former is metallic and has no 
dimerization, whereas the latter is insulating and dimerized. 

When the acceptor, $X$, is Rb$M^{\prime}$(SCN)$_4$ ($M^{\prime}=$Co, Zn), (ET)$_2X$ is a metallic 
$\theta$-type material at room temperature. When this Rb salt is cooled slowly (slow-cooling condition), 
it undergoes a structural phase transition from $\theta$ type to ${\theta}_d$ at about 
190K($\equiv T_{\text{MI}}$), and becomes insulator~\cite{3, 14, 15}. Accompanying this transition, 
the electronic state also changes: above $T_{\text{MI}}$, a metallic short-range 
charge order (CO) characterized by a diffuse rod, $\mathbf{q}=(1/3,k,1/4)$, is observed in X-ray 
measurements~\cite{4}, while below $T_{\text{MI}}$, an insulating long-range CO characterized by 
a diffuse rod, $\mathbf{q}=(0,k,1/2) \equiv {\mathbf{q}}_1$, is observed in NMR, X-ray, Raman 
scattering, and optical conductivity experiments~\cite{4, 6, 7, 8, 9, 10, 11, 12, 13}. 
The latter CO is called the horizontal CO, which has two-fold periodicity in real space. 

On the other hand, when $X$ is Cs$M^{\prime}$(SCN)$_4$ ($M^{\prime}=$Co, Zn), the crystal remains 
$\theta$ type down to low temperature and no phase transition is observed~\cite{3}. 
Above 20K, it is metallic and a short-range 3-fold diffuse rod, 
$\mathbf{q}=(2/3,k,1/3) \equiv {\mathbf{q}}_2$, is observed. 
At about $T=20$K, the resistance starts to increase by $10^{4-6}$ times accompanied by an appearance 
of the short-range diffuse rod at $\mathbf{q} = {\mathbf{q}}_1$, which coexists with 
${\mathbf{q}}_2$~\cite{16, 17}. 
The wave number, ${\mathbf{q}}_1$, is same as that observed in Rb salts below $T_{\text{MI}}$ i.e., 
the horizontal CO. 
Since the horizontal CO will be accompanied by the ${\theta}_d$ type lattice structure, Cs salts are 
supposed to be in a super-cooled glass state consisting of inhomogeneous mixture of $\theta$ and 
${\theta}_d$-type lattice structures below 20K~\cite{2, 6}. 

Recently, surprising current-voltage ($I$-$V$) characteristics are observed in Cs salts below 20K, 
where the short-range 3-fold and the horizontal CO coexist: a gigantic nonlinear conductivity~\cite{18, 19} 
and a thyristor like $I$-$V$ characteristic~\cite{21}. 
Furthermore, the X-ray measurements under a constant current below 5K show that the intensity of the 
horizontal diffuse rod decreases whereas that of the 3-fold remains. Therefore, the observed anomalous 
$I$-$V$ characteristics are assumed to be due to the different dependences of CO state on the electric 
field or current. In Rb salts, under a rapid-cooling condition, the same phenomena as Cs salts have 
been observed~\cite{21}. 

The main purpose of this paper is to understand these phenomena using an extended Hubbard 
model which is a canonical model describing charge ordered organic conductors. 
The preceding theoretical researches about charge ordering phenomena are often focused on stability 
and phase diagram under equilibrium condition~\cite{28, 29, 30, 31, 32, wo}. In this paper, 
we formulate static nonequilibrium Hartree approximation (SNHA) using Green's functions 
introduced by Keldysh~\cite{43}. 
In this method, the free energy in the presence of static electric field can be calculated. 
The $I$-$V$ characteristic is also 
investigated. We will show that the electric field stabilizes the 3-fold CO when 
the horizontal and 3-fold states are quasi degenerate. This may explain the X-ray 
measurements under a constant current. 

The model for the calculations is based on the following three hypotheses. (1) 
All calculations are carried out in the $\theta$ or ${\theta}_d$-type crystal, 
although Cs salts may have a glass state consisting of the 3-fold state with the $\theta$-type lattice and 
the horizontal state 
with the ${\theta}_d$-type lattice structure. (2) The 
metallic state under the electric field is assumed to be the 3-fold state in the ${\theta}_d$-type 
crystal in order to compare the energies, although experimentally, the crystal 
structure transforms into a new structure under the electric field, which is slightly different both from 
the $\theta$ and ${\theta}_d$-type. (3) We compare the energies of the 3-fold and horizontal 
state calculated in the uniform states. This means that the boundary energies between the 
metallic and the insulating regions are neglected. 

\begin{figure}[tb]
\begin{center}
\includegraphics[width=5cm,clip]{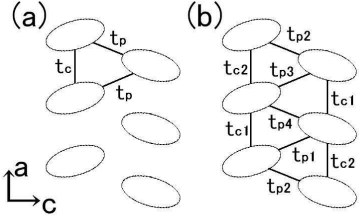}
\end{center}
\caption{Crystal structure and transfer integrals of (a) ${\theta}$ and (b) ${\theta}_d$-(ET)$_2X$ 
in a donor layer.}
\label{fig:transfer}
\end{figure}

Our Hamiltonian is written as the following extended Hubbard model: 
\begin{equation}
\begin{split}
	H = \sum_{\nnpair{x,y} \sigma} \left ( t_{xy}c^{\dagger}_{x \sigma} c_{y \sigma} 
	+ h.c. \right ) &+ U \sum_{x} n_{x \uparrow}n_{x \downarrow} \\
	&+ \sum_{\nnpair{x,y}} V_{xy} n_{x} n_{y}, \label{eq:ehh}
\end{split}
\end{equation}
where $t_{xy}$ is the transfer integral between sites, $x$ and $y$, and $U$ and $V_{xy}$ are on-site 
and nearest-neighbor Coulomb repulsion, respectively. The summation, $\sum_{\nnpair{x,y}}$, is over 
the pairs of nearest-neighbor sites. The operators, $c^{\dagger}_{x \sigma}$ ($c_{x \sigma}$) and 
$n_{x \sigma}$ represent a creation (an annihilation) and a number operator at site, 
$x$, with spin, $\sigma$. Throughout our calculations, $U$ and the ratio $V_c/U$, are fixed at 
$U=0.7$eV and $V_c/U=0.3$, respectively, which are typical values for organic 
conductors. The transfer integrals, $t_{xy}$, are $t_c = 4.3$meV and $t_p = 11.3$meV for the 
$\theta$-type crystal as shown in Fig.~\ref{fig:transfer}.~(a), and those for 
the ${\theta}_d$-type crystal are given in Fig.~\ref{fig:transfer}.~(b) and Table \ref{transferd}.
\begin{table}[tb]
\caption{Transfer integrals of ${\theta}_d$-(ET)$_2$RbZn(SCN)$_4$ at 90K calculated by the extended 
H\"uckel method~\cite{4}.}
\label{transferd}
\begin{tabular}{ccccccc}
\hline
Transfer integral & \multicolumn{6}{c}{${\theta}_d$-(ET)$_2$RbZn(SCN)$_4$} \\
 & c1 & c2 & p1 & p2 & p3 & p4 \\ \hline
($\times 10^{-2}$eV) & 1.5 & 5.2 & 16.9 & -6.5 & 2.2 & -12.3 \\
\hline
\end{tabular}
\end{table}

When we apply the Hartree approximation, the Hamiltonian, $H$, is transformed into two parts, 
$H_{\text{MF}}$ and a constant term. The former is bilinear in the electron operators. 
Taking twelve molecular sites as a unit cell, $H_{\text{MF}}$ can be 
diagonalized in momentum space as,  
\begin{equation}
	H_{\text{MF}} = \sum_{\mathbf{k} \alpha \sigma} {\tilde c}^{\dagger}_{\mathbf{k} \alpha 
	\sigma } {\varepsilon}_{\mathbf{k} \alpha \sigma} {\tilde c}_{\mathbf{k} \alpha \sigma }. 
	\label{eq:hfh}
\end{equation}
Here a new annihilation (creation) operator is written as ${\tilde c}_{\mathbf{k} \alpha \sigma} 
\equiv \sum_{\beta} u^{*}_{\mathbf{k} \beta \alpha } c_{\mathbf{k} \beta \sigma }$, 
where $\alpha$ and $\beta$ represent the band index 
and the site number in a unit cell, and $u_{\mathbf{k} \beta \alpha}$ and 
${\varepsilon}_{\mathbf{k} \alpha}$ are the eigenvector and the eigenvalue of $H_{\text{MF}}$, 
respectively. In this paper, a unit cell includes 12 molecules in order that both 3-fold and 
horizontal CO are fit in the unit cell. On the other hand, when we study the $\theta$-type 
crystal, a unit cell with 3 molecules is used, because the $\theta$-type crystal has no 
solution of the horizontal CO in the realistic parameter region. 

In the SNHA method that we formulate, observables and self-consistent equations can be derived in 
terms of the Green's functions introduced by Keldysh~\cite{43}. Here, we use the following 
Green's functions, so-called the \textit{lesser} and \textit{greater} Green's functions 
in the real space: 
\begin{equation}
\begin{cases}
	& G^{<}_{xy \sigma} (\tau , T) 
	= i \ave{{\tilde c}^{\dagger}_{y \sigma} (t^{\prime}) {\tilde c}_{x \sigma} (t)} \\
	& G^{>}_{xy \sigma} (\tau , T) 
	= -i \ave{{\tilde c}_{x \sigma} (t) {\tilde c}^{\dagger}_{y \sigma} (t^{\prime}) }, 
	\label{eq:gs}
\end{cases}
\end{equation}
with $\tau \equiv t-t^{\prime}$ and $T=\frac{1}{2}(t+t^{\prime})$, where the dependence of 
$G$ on $T$ can be ignored. In these definitions, ${\tilde c}_{x \sigma} (t)$ 
(${\tilde c}^{\dagger}_{x \sigma} (t)$) is the annihilation (creation) operator of 
${\tilde H}_{\text{MF}}$, which also includes the electric field term, 
$-e n_x (t) \mathbf{x} \cdot \mathbf{E}$, and the impurity scattering term, both of which are bilinear. 
In terms of the \textit{lesser} Green's function in Eq.~(\ref{eq:gs}), 
$\ave{{\tilde H}_{\text{MF}}}$ can be obtained as, 
\begin{equation}
\begin{split}
	\ave{{\tilde H}_{\text{MF}}} &= \lim_{\eta \to 0^+} \lim_{y \to x} \sum_{x \sigma} 
	{\left [ \frac{d}{d \tau} G^<_{xy \sigma} (\tau) \right ]}_{\tau = - \eta} \\
	& = -i \lim_{\eta \to 0 } \sum_{\mathbf{k} \alpha \sigma} 
	\int \frac{d \omega}{2 \pi} \omega e^{i \omega \eta} G^<_{\alpha \sigma}(\mathbf{k}, \omega ), 
	\label{eq:ene-g}
\end{split}
\end{equation}
where $G^<_{\alpha \sigma}(\mathbf{k}, \omega )$ is the Fourier transform of $G^<_{xy \sigma} (\tau)$. 

Then, according to the theory of quantum kinetic equation based on the Keldysh's 
method~\cite{hj}, the \textit{lesser} Green's function, 
$G^{<(1)} (\mathbf{k}, \omega)$ in the first order of $|\mathbf{E}|$ can be obtained as: 
\begin{equation}
	G^{<(1)} (\mathbf{k}, \omega) \sim i a(\mathbf{k}, \omega ) \left \{ n_{\text{F}} (\omega ) 
	- {\Delta}_{\mathbf{k}} n^{\prime}_{\text{F}} (\omega) \right \}, \label{eq:1st-order}
\end{equation}
where $n_\text{F}(\omega)$ represents the Fermi distribution function. In (\ref{eq:1st-order}), 
$a(\mathbf{k}, \omega )$ and ${\Delta}_{\mathbf{k}}$ are, 
\begin{align}
	&a(\mathbf{k}, \omega ) = \frac{\gamma }
	{ {(\omega - {\varepsilon}_{\mathbf{k} \alpha \sigma})}^2 + {(\gamma /2)}^2} \label{eq:a} \\
	&{\Delta}_{\mathbf{k}} = e \mathbf{E} \cdot \frac{{\hbar}^2 \mathbf{k}}{m} \frac{1}{\gamma}, 
	\label{eq:delta}
\end{align}
where the effective electron mass, $m$, in (ET)$_2X$ is assumed to be the free electron mass. 
In this formulation~\cite{hj}, the relaxation rate, 
$\gamma$, naturally appears in $a(\mathbf{k}, \omega )$ and ${\Delta}_{\mathbf{k}}$ due to the 
presence of impurity scattering. In our calculation, the relaxation time, $\tau = \hbar / \gamma$ 
is fixed at $10^{-8}$s in order to reproduce the resistivity. 

We have extended the calculation for $ G^{<(2)} (\mathbf{k}, \omega)$ up to the second 
order with respect to $|\mathbf{E}|$. After some straightforward manipulations, we obtain,  
\begin{equation}
	G^{<(2)}_{\alpha \sigma } (\mathbf{k}, \omega ) 
	\sim i a(\mathbf{k}, \omega ) \left \{ n_{\text{F}} (\omega ) 
	- {\Delta}_{\mathbf{k}} n^{\prime}_{\text{F}} (\omega) 
	+ { {\Delta}_{\mathbf{k}}}^2 n^{\prime \prime}_{\text{F}} (\omega) \right \}, 
	\label{eq:2nd-order}
\end{equation}
which we approximate into the following expression: 
\begin{equation}
	G^<_{\alpha \sigma } (\mathbf{k}, \omega ) \sim i a(\mathbf{k}, \omega ) 
	n_{\text{F}} (\omega - {\Delta}_{\mathbf{k}}). \label{eq:negf}
\end{equation}
Details of calculations will be published elsewhere. 
Since $\gamma$ ($\sim 10^{-7}$eV), is negligible compared to $k_BT$($\le 10^{-3}$eV), we approximate 
$a(\mathbf{k}, \omega )$ with delta function, 
$2 \pi \delta (\omega - {\varepsilon}_{\mathbf{k} \alpha \sigma})$. 
With use of the \textit{lesser} Green's function in (\ref{eq:negf}), 
$\ave{{\tilde H}_{\text{MF}}}$ becomes, 
\begin{equation}
	\ave{{\tilde H}_{\text{MF}}} = \sum_{\mathbf{k} \alpha \sigma} 
	{\varepsilon}_{\mathbf{k} \alpha \sigma} n_{\text{F}} (\omega - {\Delta}_{\mathbf{k}}). 
	\label{eq:hene}
\end{equation}
In the same way, the self-consistent equations and expectation value of free energy and 
current are obtained in terms of 
$G^<_{\alpha \sigma} (\mathbf{k} , \omega)$. 

\begin{figure}[tb]
\begin{center}
\includegraphics[width=5cm,clip]{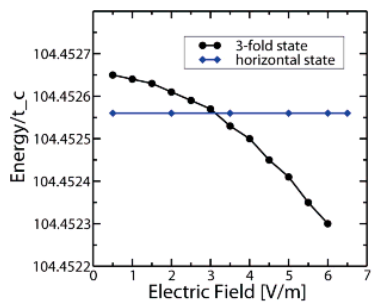}
\end{center}
\caption{Electric field dependence of the ground state energies at $V_p/V_c=0.9$.
The 3-fold CO state is stabilized by the electric field and it has a lower
energy than the horizontal state for $|{\mathbf{E}}_c|>3.2$V/m}
\label{fig:ae6}
\end{figure}

Figure \ref{fig:ae6} shows the obtained ground state energies of horizontal and 3-fold CO in the 
${\theta}_d$ type crystal for the case with $V_p/V_c=0.9$. 
If the electric field is applied more than the critical field, $E_c$, the 
3-fold state is stabilized in comparison to the horizontal state. We also calculated the gournd state 
energies for $0.8 \le V_p/V_c \le 1.0$. 
When $V_p/V_c$ is larger, the 3-fold CO state is more stable than 
the horizontal CO state.   

\begin{figure}[tb]
\begin{center}
\includegraphics[width=5cm,clip]{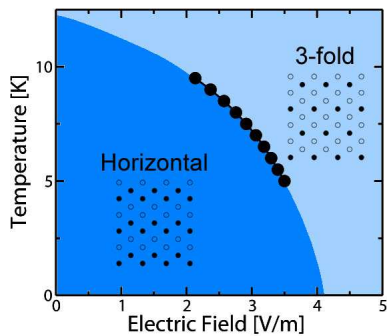}
\end{center}
\caption{The phase diagram against the temperature and the electric field.
The small solid and open circles in each region represent the charge distribution of each
CO.}
\label{fig:phase}
\end{figure}

Next, we calculate the free energies of two CO states at finite temperature, $T$, ranged from 
5K to 10K. 
The anisotropy, $V_p/V_c$, is fixed at $0.89998$ so that the energetic competition 
between the 3-fold and horizontal states is clearly seen. 
Figure \ref{fig:phase} shows the critical electric field for various 
temperatures, which gives a phase diagram. 
As $T$ increases, the 3-fold state is stabilized. 
Note that the free energy comparison is carried out in the ${\theta}_d$-type structure, although 
experimentally the metallic 3-fold state is in the $\theta$-type structure. 
In order to check the differences between the 3-fold state in $\theta$ and ${\theta}_d$-type structure, 
we have studied the free energy of the 3-fold state in the $\theta$-type, and found that the electric-field 
dependences are similar quantitatively. Therefore, if we compare the energies between the 3-fold state in 
the $\theta$-type, we expect to obtain the same phase diagram as in Fig.~\ref{fig:phase}.

We have also investigated the amplitude of the horizontal and 
3-fold CO as a function of the electric field (not shown here). In this case we studied that the 
CO in the $\theta$-type crystal instead of ${\theta}_d$ type, because the 
3-fold CO in the ${\theta}_d$ type has the horizontal modulation. 
The obtained results show that 
the electric field dependence of the amplitude of the 3-fold CO is $I=-aE^2+1.0$, with 
$a= 1.03\times 10^{-6}$ at $T=9.5$K. 
The more the electric field is applied, the weaker the intensity becomes. If the electric 
field is about 500V/m, where nonlinear conductivity is experimentally observed, 
we can expect $26$\% reduction of the amplitude, although our formulation can not be applied in such 
a strong field. On the other hand, the 
amplitude of the horizontal CO does not change by the electric field. 

\begin{figure}[tb]
\begin{center}
\includegraphics[width=6cm,clip]{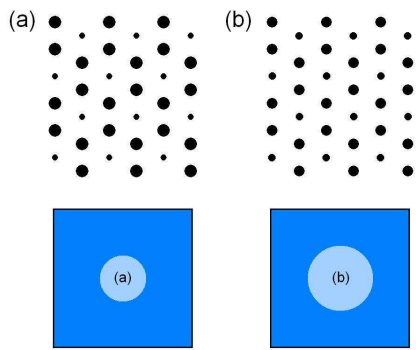}
\end{center}
\caption{Intensity, $I$, of a diffuse rod, $\mathbf{q}$, reflects both intensity (upper panels)
and area of charge modulation (lower panels).}
\label{fig:xray}
\end{figure}

These results suggest that the electric field stabilizes the 3-fold CO but decrease 
its amplitude. We expect that the experimentally observed intensity of the diffuse rod, ${\mathbf{q}}_2$, 
is affected by these two effects. 
Let us assume two cases as shown in Fig.~\ref{fig:xray}. In (a), the 
amplitude of the 3-fold CO is twice stronger than that in (b). However, 
its area is half of that of (b). If one measures the X-ray diffraction in these two cases, 
one may observe diffuse rods with roughly the same intensity. 
Although it is necessary to consider the domain size and the boundary energies, it is possible 
that the total intensity of the 3-fold CO state does not change under the 
electric field, while that of the horizontal state decreases. If this is the case, the present 
calculation explains the tendency of what are observed in the X-ray measurements 
by Sawano et al.~\cite{18}.

Finally, we would discuss the validity of the SNHA method. In the region of $|\mathbf{E}|<$1.0V/m,  
the shift ${\Delta}_{\mathbf{k}}$ in the nonequilibrium Green's function in Eq.~(\ref{eq:delta}) is 
about 0.01$t_c$, which is about 1/100 of the band gap of ${\theta}_d$-(ET)$_2$X obtained in the 
tight-binding approximation. Although ${\Delta}_{\mathbf{k}}$ is small, there is an error in the 
replacement of $G^{<(2)}(\mathbf{k}, \omega)$ in (\ref{eq:2nd-order}) into $G^< (\mathbf{k}, \omega)$ 
in (\ref{eq:negf}). This error can be estimated by,
\begin{equation}
	\delta = \left | \frac{G^{<(2)} (\mathbf{k}, \mu ) 
	- G^< (\mathbf{k}, \mu)}{G^{<(2)} (\mathbf{k}, \mu )} \right |, 
\end{equation} 
which is about $10^{-3}$ for $|\mathbf{E}|<1.0$V/m.  
However, when $|\mathbf{E}|$ is larger than 1.0V/m, the error exceeds 0.3, 
since $\beta {\Delta}_{\mathbf{k}}$ 
appearing in $n_F(\omega - {\Delta}_{\mathbf{k}})$ becomes close to 1 at the low temperature 
(5K, for example). 
This means that in the higher electric fields, the error becomes larger, and higher order correlations 
in (\ref{eq:2nd-order}) become necessary. This is the reason why we only applied the SNHA for 
$|\mathbf{E}|<6.0$V/m in Fig.~\ref{fig:ae6} and Fig.~\ref{fig:phase}. To deal with the higher electric 
fields, further improvements are necessary. 

\begin{figure}[tb]
\begin{center}
\includegraphics[width=4.5cm,clip]{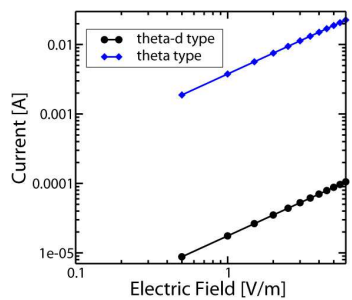}
\end{center}
\caption{$I$-$V$ characteristics of 3-fold CO in $\theta$ and ${\theta}_d$ type crystals show 
linear conductivity.}
\label{fig:lc}
\end{figure}

\begin{figure}[tb]
\begin{center}
\includegraphics[width=5cm,clip]{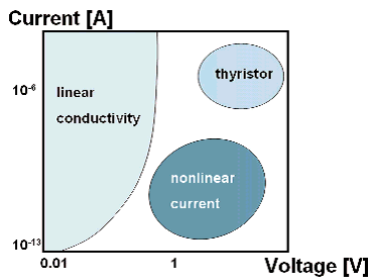}
\end{center}
\caption{Three parameter regions give linear conductivity, non-linear conductivity, and 
tylistor-like characteristic, respectively.}
\label{fig:region}
\end{figure}

When we calculate the $I$-$V$ characteristics in the present formalism, we obtain 
the linear conductivity as shown in Fig.~\ref{fig:lc}. This is reasonable because the 
region that we study corresponds to the region where the linear conductivity is observed experimentally. 
The parameter regions are shown in Fig.~\ref{fig:region} schematically. 
The present calculation corresponds to the left-hand side of this diagram. 
In this linear-conductivity region, the calculated resistivities of 
$\mathcal{O}(10^{-1})\Omega$cm in the ${\theta}_d$-type crystal 
are consistent with the measured resistivities of Rb salts of BEDT-TTF at about 200K. 
The calculated resistivities in the 3-fold CO of $\mathcal{O}(10^{-3})\Omega$cm in the $\theta$-type 
are 1/10 of the measured resistivities of Cs salts 
at about 200K. 

In summary, we formulate the static nonequilibrium Hartree approximation (SNHA) method based on 
Keldysh's Green's functions in order to calculate the observables under the constant electric field. 
By applying this method to the 3/4-filled extended Hubbard model, we study the ground state energy 
with respect to anisotropy, $V_p/V_c$, the free energy at various temperatures, charge disproportion 
in a unit cell, and $I$-$V$ characteristics.
The results show that the electric field stabilizes 
the 3-fold state in comparison to the horizontal state. The charge amplitude of the 3-fold 
CO is decreased by the applied field, while the 
horizontal CO state is not affected. 
We speculate that these results explain 
the X-ray measurement in $\theta$-(ET)$_2$X at low temperature, in which the diffuse rod, 
${\mathbf{q}}_1$, has tendency to disappear, whereas ${\mathbf{q}}_2$ remains. 
The $I$-$V$ characteristics for $|\mathbf{E}|<6.0$V/m is almost consistent with the 
experiments. Since the SNHA method cannot be applied to higher electric 
field, we did not discuss the gigantic nonlinear conductivity and thylistor-like conductivity observed 
experimentally. 
Although further improvements in the SNHA method are necessary to deal with the high-field problems, 
the present method predicts tendency of behaviors of CO states under the static electric field. 

\section*{Acknowledgment} 
We are grateful to professors H. Mori, I. Terasaki, and T. Nakamura for their fruitful 
advices. The present work is financially supported by Grands-in-Aid for Scientific Research on Priority 
Areas from MEXT, Japan.

\end{document}